\documentclass[12pt,eqsecnum,tightenlines,floats,aps,amsmath,
amssymb,nofootinbib,prd,noshowpacs,notitlepage,longbibliography]{revtex4}
\usepackage[utf8]{inputenc}
\usepackage{amsmath,amsthm,latexsym,amssymb,amsfonts,color,graphicx}
\usepackage{hyperref,epstopdf}

\global\arraycolsep=1truept
\newcommand{\beq}{\begin{equation}}
\newcommand{\eeq}{\end{equation}}
\def\bea{\begin{eqnarray}}
\def\eea{\end{eqnarray}}
\definecolor{darkred}{rgb}{.8,0,0}

\definecolor{darkblu}{rgb}{0,0,.8}

\def\pa{\partial}
\def\vs{\vspace{4mm}}
\def\ua{{\underline a}} 
\def\ub{{\underline b}} 
 
\def\uc{{\underline c}} 
 
\def\u{\underline } 
\newcommand{\BOX}[1]{
\begin{center}\fbox{\parbox{146mm}{#1}}\end{center}}

\makeatletter
\def\flE{\begin{picture}(0,0)
   \put( 0.25,    0){\vector( 1, 0){0.50}}
   \@ifstar{\@flE}{\@@flE}}
\def\@flE  #1{\put( 0.5 ,-0.03){\makebox(0,0)[ t]{$#1$}}\end{picture}}
\def\@@flE #1{\put( 0.5 , 0.03){\makebox(0,0)[ b]{$#1$}}\end{picture}}
\def\flNE{\begin{picture}(0,0)
   \put( 0.18, 0.18){\vector( 1, 1){0.64}}
   \@ifstar{\@flNE}{\@@flNE}}
\def\@flNE #1{\put( 0.52, 0.48){\makebox(0,0)[tl]{$#1$}}\end{picture}}
\def\@@flNE#1{\put( 0.48, 0.52){\makebox(0,0)[br]{$#1$}}\end{picture}}
\def\flN{\begin{picture}(0,0)
   \put(    0, 0.20){\vector( 0, 1){0.60}}
   \@ifstar{\@flN}{\@@flN}}
\def\@flN  #1{\put( 0.03, 0.5 ){\makebox(0,0)[ l]{$#1$}}\end{picture}}
\def\@@flN #1{\put(-0.03, 0.5 ){\makebox(0,0)[ r]{$#1$}}\end{picture}}
\def\flNW{\begin{picture}(0,0)
   \put(-0.18, 0.18){\vector(-1, 1){0.64}}
   \@ifstar{\@flNW}{\@@flNW}}
\def\@flNW #1{\put(-0.48, 0.52){\makebox(0,0)[bl]{$#1$}}\end{picture}}
\def\@@flNW#1{\put(-0.52, 0.48){\makebox(0,0)[tr]{$#1$}}\end{picture}}
\def\flW{\begin{picture}(0,0)
   \put(-0.25,    0){\vector(-1, 0){0.50}}
   \@ifstar{\@flW}{\@@flW}}
\def\@flW  #1{\put(-0.5 , 0.03){\makebox(0,0)[ b]{$#1$}}\end{picture}}
\def\@@flW #1{\put(-0.5 ,-0.03){\makebox(0,0)[ t]{$#1$}}\end{picture}}
\def\flSW{\begin{picture}(0,0)
   \put(-0.18,-0.18){\vector(-1,-1){0.64}}
   \@ifstar{\@flSW}{\@@flSW}}
\def\@flSW #1{\put(-0.52,-0.48){\makebox(0,0)[br]{$#1$}}\end{picture}}
\def\@@flSW#1{\put(-0.48,-0.52){\makebox(0,0)[tl]{$#1$}}\end{picture}}
\def\flS{\begin{picture}(0,0)
   \put(    0,-0.2 ){\vector( 0,-1){0.60}}
   \@ifstar{\@flS}{\@@flS}}
\def\@flS  #1{\put(-0.03,-0.5 ){\makebox(0,0)[ r]{$#1$}}\end{picture}}
\def\@@flS #1{\put( 0.03,-0.5 ){\makebox(0,0)[ l]{$#1$}}\end{picture}}
\def\flSE{\begin{picture}(0,0)
   \put( 0.18,-0.18){\vector( 1,-1){0.64}}
   \@ifstar{\@flSE}{\@@flSE}}
\def\@flSE #1{\put( 0.48,-0.52){\makebox(0,0)[tr]{$#1$}}\end{picture}}
\def\@@flSE#1{\put( 0.52,-0.48){\makebox(0,0)[bl]{$#1$}}\end{picture}}
\def\capsa(#1,#2)#3{\put(#1,#2){\makebox(0,0){$#3$}}}
\def\indiag{\@ifnextchar [{\@indiag}{\@indiag[15ex]}}
\def\@indiag[#1](#2,#3){\begingroup
   \setlength{\unitlength}{#1}
   \medskip
   \begin{center}
   \begin{picture}(#2,#3)}
\def\exdiag{\end{picture}
   \end{center}
   \medskip
   \endgroup}
\makeatother

\begin{document}





\title[]{Noether symmetries for fields and branes in 
backgrounds with Killing vectors}

\author{Josep M. Pons{$^{\dagger} $}}\email{pons@fqa.ub.edu}

\affiliation{$^{\dagger}$ Departament de F\'\i sica Qu\`antica i Astrof\'\i{s}ica and
Institut de Ci\`encies del Cosmos (ICCUB), Facultat de F\'\i sica, Universitat de Barcelona, 
Mart\'{\i} Franqu\`es 1,
08028 Barcelona, Catalonia, Spain.}

\title{Noether symmetries for fields and branes in 
backgrounds with Killing vectors}

\vskip 2truecm


\bigskip
\begin{abstract}
{\small We show that Belinfante construction of an improved energy-momentum tensor can be 
carried over to curved backgrounds, in analogy to the case of flat spacetime. The results hold 
irrespective of the background being dynamical or a fixed, non-backreacting one. It turns out 
that the analogous would-be canonical energy-momentum tensor is not covariantly conserved in general,
but 
its Belinfante ``improvement'' is. We relate this last tensor with the Hilbert tensor obtained by 
functionally derivating the Lagrangian with respect to the metric. When the background in non-dynamical, 
we discuss some issues concerning the 
Noether conserved currents associated with its Killing symmetries and the role played 
by the Belinfante tensor. Next we study extended objects ($p$-branes) either in a dynamic 
or in a fixed background, and obtain the Noether identities associated both with 
target spacetime and world volume diffeomorphisms. 
We show that in field theory as well as with extended objects, the Killing symmetries of the 
background become ordinary rigid Noether symmetries of the theory in this fixed background. 
With the example of Maxwell theory in Minkowski spacetime we show in an appendix the role of the 
Belinfante tensor in the construction of these symmetries.}
\end{abstract}


\maketitle

\vskip 1truecm
\vfill\eject

\section{Introduction}
\setcounter{equation}{0}

This paper is devoted to the formulation of classical fields and branes in curved spacetime. 
In particular we explore as to whether some results holding for flat spacetime can be extended
to curved backgrounds. In this sense this paper generalizes \cite{Pons:2009nb}, where some aspects   
of field theories with the Poincar\'e group of symmetries in flat spacetime are studied. 
In such theories it is well known that
there are essentially two equivalent procedures to define a symmetric energy-momentum tensor.
In one procedure, due to Belinfante \cite{Belinfante}, one starts with the canonical energy-momentum 
tensor obtainable form the Noether theorem \cite{Noether:1918zz} and applies to it 
an improvement in the form of a 
divergence. In the second method, due to Hilbert \cite{Hilbert}, one covariantizes the Lagrangian and 
defines the tensor as its Euler-Lagrange derivative with respect to the 
metric, setting at the end the metric to its original, Minkowski form. Rosenfeld \cite{Rosenfeld}
showed that both methods coincide on shell (in fact Belinfante tensor is only symmetric in general on 
shell).

One may wonder whether there exists some analogous construction for curved spacetime. 
In this paper we show that this construction is possible for matter Lagrangians with minimal 
couplings to gravity. An important difference is that the Riemann tensor appears as a possible 
obstruction to the conservation of the would-be canonical energy-momentum tensor, 
whereas its ``improvement'' {\sl \`a la Belinfante} overcomes this obstruction and it is always 
conserved on shell. In fact Belinfante's tensor can be defined directly form the Hilbert tensor with
no restriction to minimal coupling cases.

We will next apply the previous results to theories formulated in fixed backgrouds with Killing vectors. 
The generally covariant matter 
Lagrangian can be truncated to a fixed background and define a new Lagrangian, which is no longer 
generally 
covariant, for the matter fields. We show that the Killing symmetries of the fixed background become 
rigid Noether symmetries of the truncated Lagrangian. The associated conserved density current 
is just the contraction of the Belinfante energy-momentum tensor with such Killing vector. A sublte
point concerning the realization in phase space of these symmetries is that to
construct their generator one must use Belinfante's, not Hilbert's, energy-momentumm tensor.

\vs

The second half of the paper is devoted to the classical dynamics of extended objects in a curved 
background, which can be either dynamical or fixed. Our focus will be the Nambu Goto brane. 
We will show the contents of the Noether identities associated both with the target 
spacetime diffeomorphisms and world volume diffeomorphism. Interestingly enough, 
we are able to obtain, just from the Noether analysis, the equations of motion (EOM) 
for the NG Lagrangian, 
that is, the vanishing of the trace of the extrinsic curvature, without doing any explicit 
computation of the Euler-Lagrange derivatives of the NG Lagrangian. 

In the case of a fixed background, we show that, in a way similar to the result for 
field theories, the Killing symmetries of the background become rigid symmetries of the 
truncated theory. A simple geometric interpretation of the associated conserved currents is also given.

\vs

The paper is organized as follows. Afer some basic preliminaries, section \ref{NI} is devoted to 
the Noether identities in a general background and the connection between Hilbert, 
Belinfante energy-momentum tensors and what we call the canonical tensor. 
In section \ref{fixkill} we discuss the fate
of the Killing symmetries in theories truncated to a fixed metric background. Next in section 
\ref{theNG} we turn to branes in the different settings of fixed or dynamic backgrounds and we 
identify for the brane Lagrangian the symmetries  associated with the Killing vectors. In section 
\ref{NIb} the Noether identities associated with target spacetime and/or world volume diffeomorphisms 
are obtained. Conclusions are drawn in section \ref{concl}. We finish with two appendices, 
the first to underline the relevance of Belinfante tensor in the construction of the generator 
of the Noether symmetries originated from Killing vectors, and the second, 
in order to make the paper more selfcontained, with a review of basic results 
on the extrinsic curvature, useful for the sections on extended objects.

\section{Preliminaries}
\setcounter{equation}{0}

Here we introduce some notation and basic observations. 
\subsection{ Infinitesimal diffeomorphisms. Praise for the active view}
Let us examine the transformation of the fields under reparametrizations of a manifold $P$, 
$x^\mu\to x'^\mu=x^\mu-\epsilon^\mu(x)$, with $\epsilon^\mu(x)$ an arbitrary infinitesimal function 
(though the infinitesimal parameter can always be factored out from $\epsilon^\mu(x)$).

Reparametrizations -that is, changes of coordinates- are the passive interpretation of diffeomorphisms.
To discus the active versus passive views of diffeomorphisms, 
consider for simplicity the case of a scalar field $\varphi(x)$. According to the passive view, 
under an infinitesimal diffeomorphism, the coordinates undergo a change, $x^\mu\to x'^\mu(x)$, 
and the fields remain the same, the only change being in their mathematical description because 
we must write them -or their components- in the new coordinatization. For the scalar 
field this is given by $\varphi'(x')=\varphi(x)$. Instead, in the active view, 
the coordinates do not change but the fields undergo the change 
$\varphi(x)\to\varphi'(x)$, so that the functional variation of $\varphi(x)$ is 
$\delta \varphi=\varphi'(x)-\varphi(x) = {\cal L}_{\!{}_\epsilon}\varphi$, where the differential 
operator ${\cal L}_{\!{}_\epsilon}$ is the Lie derivative under the vector field $\epsilon^\mu(x)$. 
In general, any vector, tensor, form, will experience an active variation given by its Lie derivative
under $\epsilon^\mu(x)$.

Some praise is deserved for the active view of spacetime symmetries, in which we only consider the 
functional variation of the fields. {\it First}, it puts the spacetime symmetries on an equal footing with 
respect to the other, internal symmetries, that may eventually exist; {\it second}, the functional variation 
$\delta$ commutes with 
the partial derivative with respect to the coordinates, $[\delta ,\, \pa_\mu ]=0$; {\it third}, 
the functional variation of an action does not modify the boundaries -because the coordinates 
are unchanged-, thus simplifying intermediate computations; {\it fourth}, if the variations are 
obtained in a canonical -phase space- formalism with a 
generator acting through the Poisson bracket, they are automatically functional variations, that is, 
of the active type. This is the case of conserved quantities acting as generators of symmetries 
according to Noether's theorem. 

When the dynamics of extended objects embedded in the manifold $P$ is considered, 
the same ideas apply. We must distiguish in this case target
spacetime diffeomorphisms and world volume diffeomorphisms, but in both of them we will apply 
the active view. 

It is this {\it active view} of diffeomorphisms that we will adopt throughout the paper: 
{\it move the structures -fields, branes-, not the coordinates}. 

\subsection{Truncation to a fixed background}
We consider theories formulated in manifolds endowed with a Lorentzian metric.
With $g$ representing the metric field and $\phi$ the matter fields, 
we will consider a matter Lagrangian ${\cal L}[g,\phi]$ fully covariant, that is, behaving as a  
scalar density under diffeomorphisms. At any moment ${\cal L}$ can be truncated to 
${\cal L}^{\!{}^{(0)}}[\phi]:= {\cal L}[g,\phi]\vert_{g\to g_0}$, where $g_0$ is a fixed 
background\footnote{By fixed background we mean a non-dynamical metric field with 
no equations of motion for it.}. 
This type of truncation is not a gauge fixing nor it is a consistent truncation. 
It is not a gauge fixing because evidently there is no diffeomorphism connecting a generic 
metric configuration to $g_0$. 
As a truncation, it is not consistent because gravity universally couples with matter and 
therefore the backreaction must not be neglected. But it may be a useful 
approximation, as it is the case of field theory in flat spacetime. 

Once the background is fixed (for instance to Minkowski spcetime) 
the gauge symmetry of diffeomorphisms is no longer there. In the Minkowski case the ``residual'' 
(not exaclty so, because fixing the background is {\sl not} a gauge fixing) symmetry is rigid: 
Poincar\'e. It is the symmetry generated by the Killing vectors of the background.
In the active interpretation one can still say that the spacetime confi\-gu\-rations, 
before and after the action of a Poincar\'e symmetry, are physically indistinguishable 
(the language remains of gauge equivalent configurations) and in fact in the passive view we see 
that we are describing a unique physical configuration with different coordinates. But now the 
indeterminism with respect to the initial conditions, which is a characterisitic feature of the 
gauge freedom, has disappeared.

\section{Energy-momentum tensors and Noether identities in field theory}
\setcounter{equation}{0}
\label{NI}

\subsection{Noether identities}
Consider a first order Lagrangian ${\cal L}[\psi]$ with gauge freedom described by 
\beq
\delta \psi^{A} = R^A_a \epsilon^a + Q^{A\mu}_a \partial_\mu\epsilon^a \ \Rightarrow \, 
\delta {\cal L} = {\rm divergence}.
\label{gaugetr}
\eeq
where $\epsilon^a$ are the infinitesimal arbitrary functions of the gauge 
symmetries (in a certain number given by the running of the index $a$).
The associated Noether identities are\footnote{This subject is quite standard, all details are given 
for instance in \cite{Pons:2009nb}.}
\beq
[{\cal L}]_{\!{}_A} R^A_a-\partial_\mu([{\cal L}]_{\!{}_A} Q^{A\mu}_a)=0\,,
\label{noethid0}
\eeq
(with the usual notation 
$\displaystyle
[{\cal L}]_{\!{}_A} = \frac{\delta {\cal L}}{\delta \psi^A}$ for the Euler-Lagrange
functional derivatives with respect to the fields -or field components- $\psi^A$) or 
equivalently, after saturating with the arbitrary functions $\epsilon^a$,
\beq
[{\cal L}]_{\!{}_A} \delta\psi^A-\partial_\mu([{\cal L}]_{\!{}_A} Q^{A\mu}_a\epsilon^a)=0\,.
\label{noethid1}
\eeq
We are interested in a special type of gauge symmetries: spacetime diffeomorphisms. 
In view of that it is convenient to express these identities
in covariant language\footnote{The only connection we consider throughout the paper is
the Levi Civita connection.}. For notational convenience, let us distinguish, among the generic 
fields $\psi^A$, the metric field 
$g_{\mu\nu}$ and the matter fields $\phi^A$, which we consider bosonic 
(fermionic matter has to be formulated within the tetrad formalism). 
We assume that ${\cal L}[g,\phi]$ depends
at most on the first spacetime derivatives of the fields $\phi^A$.

As usual in the framework of generally covariant theories, we consider that 
the matter Lagrangian ${\cal L}[g,\phi]$ behaves as a scalar density under diffeomporphisms. 
Thus diffeomporphisms are Noether gauge symmetries for the theory described by this Lagrangian.
In the active view, diffeomorphism covariance is expressed infinitesimally with the transformations 
(Lie derivative)
\beq\delta g_{\mu\nu} = \nabla_{\!\mu}\, \epsilon_\nu + 
\nabla_{\!\nu}\, \epsilon_\mu\,,\qquad
\delta \phi^A = \epsilon^\rho \nabla_{\!\rho} \phi^A + 
Q^{A\,\rho}_{\quad\, \sigma}\nabla_{\!\rho}  \epsilon^\sigma
\label{deltas}
\eeq
(Indices $A$ for fields or field components may generically include spacetime indices as well 
as internal indices), which are just the Lie derivatives of the metric and the matter fields 
under an infinitesimal spacetime vector field $\epsilon^\mu$. The covariant derivative is 
defined with the Levi Civita connection. These transformations (\ref{deltas}) will induce 
$\displaystyle\delta {\cal L}= \partial_\mu(\epsilon^\mu{\cal L})$. 
The Noether identities (\ref{noethid1}), now associated with diffeomorphism covariance, read
(with $\displaystyle [{\cal L}]^{\mu\nu}= \frac{\delta {\cal L}}{\delta g_{\mu\nu}}$)
\beq
[{\cal L}]^{\mu\nu}\delta g_{\mu\nu} + [{\cal L}]_{\!{}_A} \delta \phi^A - 
2 \nabla_{\!\lambda}([{\cal L}]^{\lambda\mu}\epsilon_\mu) -
\nabla_{\!\lambda}([{\cal L}]_{\!{}_A} Q^{A\,\lambda}_{\quad\, \mu}\,  \epsilon^\mu)=0\,,
\label{noethid2}
\eeq
identically\footnote{We have replaced the covariant derivative for the partial derivative because
they coincide when defining the divergence of a vector density of weight one. This is also true for 
the divergence of an antisymmetric tensor density. This replacement will be made at convenience 
in the following sections with no further warning.}.

\subsection{Hilbert and Belinfante's energy-momentum tensors}

The {\sl Hilbert energy-momentum tensor density} is defined as 
$\displaystyle 
T^{\mu\nu} := -2 \frac{\delta {\cal L}}{\delta g_{\mu\nu}}\,, 
$ and so we can write
\beq 
-T^{\mu\nu} \nabla_{\!\mu}\, \epsilon_\nu + [{\cal L}]_{\!{}_A}\delta \phi^A + 
\nabla_{\!\lambda} \Big( (T^{\lambda}_{\ \mu} - 
[{\cal L}]_{\!{}_A} Q^{A\,\lambda}_{\quad\, \mu}) \epsilon^\mu  \Big) =0   \,,
\label{noethid3}
\eeq
for arbitrary $\epsilon^\mu$. Comparing with the standard results in the 
Minkowski case (see details in \cite{Pons:2009nb}), it is reasonable to define the combination 
\beq
T^{\lambda}_{\! \ub\ \mu}:=T^{\lambda}_{\ \mu} - 
[{\cal L}]_{\!{}_A} Q^{A\,\lambda}_{\quad\, \mu}
\label{belinf}
\eeq as the {\sl Belinfante energy-momentum tensor density}\footnote{We underline, $\ub$, to distinguish
the notation for the Belinfante tensor $T_{\! \ub }$ from a tensorial index. 
We will do the same for the canonical tensor below.}, 
which will connected later -see next subsection- be with the canonical tensor.

Defining $\displaystyle Q^A_\mu  =  R^A_\mu - Q^{A\nu}_{\quad\,\rho} \Gamma^\rho_{\mu\nu}$, with 
$\Gamma^\rho_{\mu\nu}$ the Christoffel symbols,  
we may write (\ref{gaugetr}) in a covariant form, 
$\displaystyle
\delta \phi^A = Q^A_\mu \epsilon^\mu + Q^{A\mu}_{\quad\,\nu} \nabla_{\!\mu}\epsilon^\nu\,.$ 
Then, integrating (\ref{noethid3}) with 
the arbitrary $\epsilon^\mu$ taken with compact support we can eliminate these arbitrary 
functions and obtain the covariant Noether identities
\beq
\nabla_{\!\mu} T^{\mu}_{\ \nu} + [{\cal L}]_{\!{}_A} Q^A_{\nu} -
\nabla_{\!\lambda}([{\cal L}]_{\!{}_A}Q^{A\lambda}_{\quad\,\nu})=0\,.
\label{noethid4}
\eeq
Notice that Eq.(\ref{noethid4}) can be understood 
as an identity satisfied by the matter Lagrangian in a fixed background.  
Clearly, since to compute its Hilbert tensor one has to functionally derivate with respect 
to the components of the metric field, one must know the Lagrangian for metrics around the 
fixed configuration, but this is our case because our starting point was a fully covariant 
scalar density Lagrangian. We observe that the Hilbert tensor -and Belinfante's, 
according to (\ref{belinf})- is covariantly conserved as long as the EOM for the matter fields are 
satisfied, irrespective of being either in a fixed or a dynamical metric background. 
More on dynamics on a fixed background in section \ref{fixkill}.

\subsection{Contact with the canonical tensor}

To continue we will asume that the matter Lagrangian depends only up to the first 
spacetime derivatives of the metric. This means that we are excluding some non-minimal 
couplings that may use the Riemann tensor\footnote{This restriction does not apply to 
the case of Killing symmetries, analyzed in the next section.}. 

In addition to Hilbert's and Belinfante's tensors, in Minkowski spacetime one defines 
the ca\-no\-nical energy-momentum tensor, 
$\displaystyle \frac{\partial {\cal L}}{\partial\, \partial_\sigma\phi^A} 
\partial_\rho \phi^A - \delta^\sigma_\rho{\cal L}$.  
In this spirit we now define the canonical tensor\footnote{Denominating this tensor the 
ca\-no\-nical energy-momentum tensor would be misleading because, as we will see below, it is not 
covariantly conserved in general.} in curved spacetime
\beq
T^{\sigma}_{\! \uc\ \rho}:= \frac{\partial {\cal L}}
{\partial\, \nabla_{\!\sigma}\phi^A} \nabla_{\!\rho} \phi^A -
\delta^\sigma_\rho{\cal L}\,.
\label{canEM}
\eeq

In the following we will connect this tensor with Belinfante's. To do so, we revisit the result 
$\displaystyle \delta_{\epsilon} {\cal L} = \partial_\mu (\epsilon^\mu{\cal L})$ 
for the Lie derivative of the Lagrangian under an arbitrary spacetime vector $\epsilon^\mu$. 
It may be written\footnote{Notice that we take the Euler-Lagrange derivative $[{\cal L}]_{\!{}_A}$ for 
$L(\phi^A,\nabla_{\!\mu}\phi^A)$
as $\displaystyle [{\cal L}]_{\!{}_A} =\frac{\partial {\cal L}}
{\partial\, \phi^A} - \nabla_{\!\mu}\frac{\partial {\cal L}}
{\partial\, \nabla_{\!\mu}\phi^A}$, so that both terms, $\displaystyle \frac{\partial {\cal L}}
{\partial\, \phi^A}$ and $\displaystyle \nabla_{\!\mu}\frac{\partial {\cal L}}
{\partial\, \nabla_{\!\mu}\phi^A}$, are geometric objects.} as
\beq
\delta_{\epsilon} {\cal L} =
\frac{\delta {\cal L}}{\delta g_{\mu\nu}}\delta g_{\mu\nu} + 
\partial_\sigma(\frac{\partial {\cal L}}{\partial g_{\mu\nu,\sigma}} \delta g_{\mu\nu}) + 
[{\cal L}]_{\!{}_A}\delta \phi^A +
\nabla_{\!\mu}(\frac{\partial {\cal L}}{\partial\, \nabla_{\!\mu}\phi^A}\delta \phi^A)=
\nabla_{\!\mu} (\epsilon^\mu{\cal L})\,.
\label{deltalag}
\eeq
As consequence of our assumption, the dependence of ${\cal L}$ on the derivatives of the 
metric will be concealed within the covariant derivatives of the fields $\phi^A$, so that
$$
\displaystyle \frac{\partial {\cal L}}{\partial g_{\mu\nu,\sigma}} = 
\frac{\partial {\cal L}}{\partial(\nabla_{\!\rho}\phi^A)}\,
\frac{\partial (\nabla_{\!\rho}\phi^A)}{\partial \Gamma^\alpha_{\beta\gamma}} \,
\frac{\partial \Gamma^\alpha_{\beta\gamma}}{\partial g_{\mu\nu,\sigma} } =:
\frac{1}{2} M^{\sigma\mu\nu}\,,
$$
with\footnote{We use the standard notation for symmetrization 
$\displaystyle N^{(ab)} = \frac{1}{2}(N^{ab}+N^{ba})$ and antisymmetrization 
$\displaystyle N^{[ab]} = \frac{1}{2}(N^{ab}-N^{ba})$ and so on.} 
$M^{\sigma\mu\nu} = M^{\sigma(\mu\nu)}$ being a tensor density because the first factor 
$\displaystyle \frac{\partial {\cal L}}{\partial(\nabla_{\!\rho}\phi^A)}$ 
in the expression above is a tensor density whereas the other two are tensors. 
Thus we can write (\ref{deltalag}) as
\beq
-T^{\mu\nu} \nabla_{\!\mu}\, \epsilon_\nu
+ \nabla_{\!\sigma}(M^{\sigma\mu\nu}\nabla_{\!\mu}\, \epsilon_\nu ) + 
[{\cal L}]_{\!{}_A}\delta \phi^A +
\nabla_{\!\mu}(\frac{\partial {\cal L}}
{\partial\, \nabla_{\!\mu}\phi^A}\delta \phi^A-\epsilon^\mu{\cal L})=0\,.
\label{deltalag2}
\eeq
Subtraction of (\ref{deltalag2}) from (\ref{noethid3}) and the use of (\ref{deltas}) 
together with definiton (\ref{canEM}) yields the -identical- conservation law 
\bea
&&\nabla_{\!\sigma} \Big( (T^{\sigma}_{\ \mu} - 
[{\cal L}]_{\!{}_A} Q^{A\,\sigma}_{\quad\, \mu}) \epsilon^\mu  - 
M^{\sigma\nu\mu} \nabla_{\!\nu}\, \epsilon_\mu -   
\frac{\partial {\cal L}}{\partial\, \nabla_{\!\sigma}\phi^A}
(\epsilon^\rho \nabla_{\!\rho} \phi^A + 
Q^{A\,\rho}_{\quad\, \lambda}\nabla_{\!\rho}  \epsilon^\lambda)+
\epsilon^\sigma{\cal L}  \Big) 
\nonumber \\
&& =
\nabla_{\!\sigma} \Big( (T^{\sigma\mu}_{\! \ub}-T^{\sigma\mu}_{\! \uc})\, \epsilon_\mu  - 
(M^{\sigma\nu\mu}+\frac{\partial {\cal L}}{\partial\, \nabla_{\!\sigma}\phi^A} 
Q^{A\,\nu}_{\quad\, \lambda}\,g^{\lambda\mu}\,) \nabla_{\!\nu}\, \epsilon_\mu  
  \Big)=0\,.
\label{conservlaw}
\eea
We will get interesting information from the fact that this conservation law, 
consequence of general covariance, holds for any spacetime 
vector $\epsilon^\mu$. Relation (\ref{conservlaw}) has the form
\beq 
A^\mu \epsilon_\mu + B^{\nu \mu} \nabla_{\!\nu}\epsilon_\mu 
+ C^{\sigma\nu\mu} \nabla_{\!\sigma}\nabla_{\!\nu}\epsilon_\mu =0\,,
\label{conservlaw2}
\eeq
identically. Being $\epsilon^\mu$ arbitary, the coefficient of 
$\displaystyle \partial_\sigma\partial_\nu\epsilon^\mu$ must vanish, which implies that 
$\displaystyle C^{\sigma\nu\mu}$ is antisymmetric in its first two upper indices:
$\displaystyle C^{\sigma\nu\mu} = C^{[\sigma\nu]\mu}\,, $ which in turn 
means that the last 
term in (\ref{conservlaw2}) is\footnote{We use the conventions in \cite{Wald:1984rg}.}
$\displaystyle C^{[\sigma\nu]\mu} \nabla_{[\sigma}\nabla{_\nu]}\epsilon_\mu
= \frac{1}{2}C^{[\sigma\nu]\mu} R_{\sigma\nu\mu}^{\quad\ \rho} \epsilon_\rho$. 
So finally we obtain
$$
C^{\sigma\nu\mu} = C^{[\sigma\nu] \mu}\,,\ B^{\nu\mu} =0\,,\ 
A_\rho + \frac{1}{2}C^{[\sigma\nu]\mu} R_{\sigma\nu\mu}^{\quad\ \rho} \epsilon_\rho=0\,,
$$
which, translated to (\ref{conservlaw}), becomes
\bea 
 C^{[\sigma\nu]\mu} &=& - M^{\sigma(\nu\mu)} - 
 \frac{\partial {\cal L}}{\partial\, \nabla_{\!\sigma}\phi^A} 
 Q^{A\,\nu}_{\quad\, \lambda}\, g^{\lambda\mu}\,,\nonumber \\
 0&=&T^{\sigma\mu}_{\! \ub}-T^{\sigma\mu}_{\! \uc} + 
 \nabla_{\!\lambda} C^{[\lambda\sigma]\mu}\,,\nonumber \\  
 0&=& \nabla_{\!\sigma}(T^{\sigma\rho}_{\! \ub}-T^{\sigma\rho}_{\! \uc}) + 
 \frac{1}{2}C^{[\sigma\nu]\mu} R_{\sigma\nu\mu}^{\quad\ \rho}\,.
\label{results}
\eea
The third equation in (\ref{results}) is 
just a consistency check for the second. The second equation gives the relation
\beq
T^{\sigma\mu}_{\! \ub}=T^{\sigma\mu}_{\! \uc} -
 \nabla_{\!\lambda} C^{[\lambda\sigma]\mu}\,.
\label{tbelinf}
\eeq
The first equation is a statement on the decomposition of 
$\displaystyle \frac{\partial {\cal L}}{\partial\, \nabla_{\!\sigma}\phi^A}
Q^{A\,\nu}_{\quad\, \lambda}\, g^{\lambda\mu}$ into partially symme\-tric and 
partially antisymmetric parts, that is
$$
-\frac{\partial {\cal L}}{\partial\, \nabla_{\!\sigma}\phi^A} 
 Q^{A\,\nu}_{\quad\, \lambda}\, g^{\lambda\mu} = C^{[\sigma\nu]\mu}+M^{\sigma(\nu\mu)}\,.
$$
As a matter of fact it is well known that given an object with three indices, 
$\displaystyle O^{\sigma\nu\mu}$, it has a unique decomposition of the form 
$\displaystyle O^{\sigma\nu\mu} = O_{\!a}^{[\sigma\nu]\mu}+O_{\!s}^{\sigma(\nu\mu)} $. 
In particular, defining the antisymmetric combination
$$
S^{A [\mu\nu]}_B \phi^B := Q^{A\,\nu}_{\quad\, \lambda}\, g^{\lambda\mu}-
Q^{A\,\mu}_{\quad\, \lambda} g^{\lambda\nu}\,,
$$
($Q^{A\,\nu}_{\quad\, \lambda}$ is linear in the fields for the Lie derivative (\ref{deltas}))
it turns out that 
\beq 
C^{[\sigma\nu]\mu} = \frac{1}{2}\Big(
\frac{\partial {\cal L}}{\partial\, \nabla_{\!\mu}\phi^A} S^{A [\nu\sigma]}_B +
\frac{\partial {\cal L}}{\partial\, \nabla_{\!\nu}\phi^A} S^{A [\mu\sigma]}_B +
\frac{\partial {\cal L}}{\partial\, \nabla_{\!\sigma}\phi^A} S^{A [\nu\mu]}_B 
\Big)\phi^B\,.
\label{Cdet}
\eeq

Expressions (\ref{tbelinf}) and (\ref{Cdet}) generalize to curved spacetime the classical formulas for
obtaining Belinfante's tensor as an improvement of the canonical tensor
(see section 4.1 of \cite{Pons:2009nb}). With a caveat: 
in curved spacetime, since (\ref{noethid4}) and (\ref{belinf}) imply that 
Belinfante's tensor is covariantly conserved on shell, we infer that  
the canonical tensor is {\sl not} conserved in general. 
Indeed, from the third equation in (\ref{results}) we obtain
\beq\nabla_{\!\sigma}T^{\sigma\rho}_{\! \uc} 
\mathrel{\mathop{=}\limits_{\hbox{\rm \tiny (on\ shell)}}} 
 \frac{1}{2}C^{[\sigma\nu]\mu} R_{\sigma\nu\mu}^{\quad\ \rho}\,,
 \label{can-not}
\eeq
with $R_{\sigma\nu\mu}^{\quad\ \rho}$ the Riemann tensor. 
Thus the construction of the Belinfante tensor out of the canonical tensor is not exactly 
an improvement but a procedure to obtain a conserved tensor which coincides on shell with Hilbert tensor 
(see (\ref{belinf})) and thus is symmetric on shell.

Only for Minkowski spacetime -with vanishing Riemann tensor- or for theories with only 
scalar fields -for which $Q^{A\,\rho}_{\quad\, \sigma}$ in (\ref{deltas}) vanishes-, 
the covariant conservation of the canonical tensor can be generally asserted.
\section{Fixed backgrounds with Killing vectors}
\setcounter{equation}{0}
\label{fixkill}
Complementing what has been already said, we notice that the basic equations of the previous section,
(\ref{noethid3}), 
(\ref{noethid4}), (\ref{deltalag2}), (\ref{conservlaw}), can be understood 
as identities satisfied by the matter Lagrangian ${\cal L}^{\!{}^{(0)}}$ truncated to a fixed 
background $g_0$.  In this section we will interpret these equations in this sense. 
As identities, they are a gift from the former diffeomorphism covariance enjoyed by the theory 
before truncation. 

An immediate consequence of identity (\ref{noethid3}) and definition (\ref{belinf}) is that if 
$\epsilon^\mu$ is a Killing (K.) vector of the fixed background, that is, 
\beq
\delta g_{\mu\nu} = \nabla_{\!\mu}\, \epsilon_\nu + 
\nabla_{\!\nu}\, \epsilon_\mu=0\,,\quad (\Leftrightarrow\ \epsilon^\mu\in K.)\,,
\label{killingcond}
\eeq
then
\beq 
 [{\cal L}^{\!{}^{(0)}}]_{\!{}_A}\delta \phi^A + 
\nabla_{\!\lambda} (T^{\lambda}_{\! \ub\ \mu} \epsilon^\mu) =0   \,,\quad (\epsilon^\mu\in K.)\,, 
\label{noethkill}
\eeq
identically. Eq.(\ref{noethkill}) neatly displays the result that the Killing symmetries 
of the background $g_0$ have morphed into rigid Noether symmetries for the 
truncated Lagrangian ${\cal L}^{\!{}^{(0)}}$. Also this equation identifies the vector density 
\beq 
J^{\lambda}:= T^{\lambda}_{\! \ub\ \mu} \epsilon^\mu  \,,\quad (\epsilon^\mu\in K.)\,,
\label{killcurrent}
\eeq
as the Noether current for this rigid symmetry, $ [{\cal L}^{\!{}^{(0)}}]_{\!{}_A}\delta \phi^A + 
\partial_{\lambda} J^{\lambda} =0$. We may say:

\vs

\BOX{\vfill\vspace{1.5mm}\hspace{1.5mm}\it In field theory, the conserved density current of a 
Noether symmetry generated by a 
Killing vector of the fixed background, 
is the contraction of the Belinfante energy-momentum tensor with such Killing vector.
\vfill\vspace{1.5mm}}

\vs

One finds in the literature a trivial proof of the existence of an the shell conserved current 
$T^{\lambda}_{\! \ \mu} \epsilon^\mu$ based on the Killing condition (\ref{killingcond}) and 
the fact that Hilbert tensor is -on shell- covariantly 
conserved, but the status of a Noether symmetry 
is more than that, and this conservation does not make the current 
to be a Noether current -although both currents coincide on shell. 
The true Noether current associated with the Killing symmetries of a 
fixed background is (\ref{killcurrent}), which uses the Belinfante -not Hilbert's- tensor. 
Switching to canonical variables, one would find as generator -under the Poisson bracket- of 
the symmetry the quantity $G:=\int d^3 x J^0$, which, under suitable conditions at the spatial boundary,
is a conserved charge. 

If the matter theory is a gauge theory there may be obstructions \cite{Pons:1999az} for this quantity 
$G$ the generate the symmetry transformations in phase space. As an illustration we work out in 
Apendix \ref{apa} some details of the application of this analysis to vacuum Maxwell theory in flat 
spacetime, and we will show the crucial role played by Belinfante tensor in constructing 
the internal gauge symmetry generators.  

\vspace{4mm}

Notice that in the case of a fixed background with Killing vectors, as far as the 
rigid Noether symmetries generated by the Killing vectors are concerned, we do not need
to restrict ourselves to the minimal coupling case leading to (\ref{deltalag}), 
because $\delta g_{\mu\nu}$ vanishes. Thus we can write (\ref{deltalag}), with no restrictions 
on the coupling to gravity, as
\beq 
\delta {\cal L}^{\!{}^{(0)}}= [{\cal L}^{\!{}^{(0)}}]_{\!{}_A}\delta \phi^A +
\nabla_{\!\mu}(\frac{\partial {\cal L}^{\!{}^{(0)}}}{\partial\, 
\nabla_{\!\mu}\phi^A}\delta \phi^A)=
\nabla_{\!\mu} (\epsilon^\mu{\cal L}^{\!{}^{(0)}}) \,,\quad (\epsilon^\mu\in K.)\,.
\label{deltalagkill}
\eeq
for the Lagrangian ${\cal L}^{\!{}^{(0)}}$ in the fixed background $g_0$ and with 
$\epsilon^\mu$ being a Killing vector of $g_0$.

Comparison of (\ref{deltalagkill}) with (\ref{noethkill}) shows the existence of an 
identically conserved density current
\beq
\nabla_{\!\sigma} \Big( T^{\sigma}_{\! \ub\ \mu} \epsilon^\mu - 
\frac{\partial {\cal L}^{\!{}^{(0)}}}{\partial\, \nabla_{\!\sigma}\phi^A} \delta \phi^A
+ \epsilon^\sigma  {\cal L}^{\!{}^{(0)}}
  \Big)= 0 \,,\quad (\epsilon^\mu\in K.)\,,
\label{conservlawkill}
\eeq
which, again, is a gift from the general covariance properties held by the theory 
before being truncated to a fixed background. 

Some comments are in order. Eq.(\ref{conservlawkill}) and Eq. Eq.(\ref{noethkill}) show 
that there are two equivalent presentations of the Noether conserved density current 
associated with 
the Killing symmetry. The ``classical'' one would have been, by typical use of Noetherian methods, 
from (\ref{deltalagkill}),
$$
J^\mu_{\!\rm \small class.} = 
\frac{\partial {\cal L}^{\!{}^{(0)}}}{\partial\, \nabla_{\!\sigma}\phi^A} \delta \phi^A
-  \epsilon^\sigma  {\cal L}^{\!{}^{(0)}} \,,\quad (\epsilon^\mu\in K.)\,,
$$
whereas the alternative presentation is (\ref{killcurrent}). 

As far as we know, little attention
is paid in the literature to the relationship existing between these two currents. 
Both currrents differ by an 
identically conserved density current which must have locally the form  
$\displaystyle J^{\mu} -J^\mu_{\!\rm \small class.} = \partial_\nu N^{\mu\nu}$, with $N^{\mu\nu}$
an antisymmetric tensor density\footnote{With a torsionless connection $\partial_\nu N^{\mu\nu}=
\nabla_\nu N^{\mu\nu} $. Notice that $\nabla_{\!\mu}\partial_\nu N^{\mu\nu}
= \nabla_{\!\mu}\nabla_{\!\nu} N^{\mu\nu} = \frac{1}{2} [\nabla_{\!\mu},\,\nabla_{\!\nu}]N^{\mu\nu}=
-\frac{1}{2} R_{\mu\nu\alpha}^{\quad\ \mu}N^{\alpha\nu}
-\frac{1}{2} R_{\mu\nu\alpha}^{\quad\ \nu}N^{\mu\alpha}=
\frac{1}{2}R_{\nu\alpha}N^{\alpha\nu}-\frac{1}{2}R_{\mu\alpha}N^{\mu\alpha}= -R_{\mu\alpha}N^{\mu\alpha}=0
$, with $R_{\mu\alpha}$ the Ricci tensor, which is symmetric.}. Thus, both 
$G= \int d^3 x J^{0}$ or $G_{\!\rm \small class.}= \int d^3 x J^{0}_{\!\rm \small class.}$ generate 
the same transformations because they differ by a boundary term.
\vspace{4mm}

\u{\bf An example}

Just as an example, let us consider 
the Lagrangian 
$${\cal L}^{\!{}^{(0)}} = \frac{1}{2} \sqrt{-g} R\, \phi^2
$$
which describes a non-minimal coupling of a scalar field with the metric through the curvature scalar, 
here for a fixed background, and suppose that $\epsilon^\mu$ is a Killing vector for this background. 
One can write (\ref{conservlawkill}) for this case as (here Hilbert and Belinfante tensors coincide)
\beq
\nabla_{\!\sigma} \Big( T^{\sigma}_{\! \ \mu} \epsilon^\mu 
+ \epsilon^\sigma  {\cal L}^{\!{}^{(0)}}
  \Big)= \nabla_{\!\sigma}(\sqrt{-g}\Big( (\nabla^\mu\nabla^\nu \phi^2) \epsilon_\nu - 
  R^{\mu\nu}  \epsilon_\nu - ( \triangle  \phi^2 ) \epsilon^\mu                        \Big)) =0 
  \,,\quad (\epsilon^\mu\in K.)\,,
\label{conservlawkillex}
\eeq
and using for the Killing vector $\epsilon^\mu$ the relation 
$R^{\mu\nu}  \epsilon_\nu =-\triangle \epsilon^\mu\quad (\epsilon^\mu\in K.)\,,$ one easily finds, 
for the relevant term in (\ref{conservlawkillex}),
$$
\sqrt{-g}\Big( (\nabla^\mu\nabla^\nu \phi^2) \epsilon_\nu - R^{\mu\nu}  \epsilon_\nu - 
  ( \triangle  \phi^2 ) \epsilon^\mu \Big) = \nabla_\nu N^{\mu\nu}= \partial_\nu N^{\mu\nu}\,,
  \quad (\epsilon^\mu\in K.)\,,
$$
with $N^{\mu\nu}$ the antisymmetric tensor density
$$N^{\mu\nu} = \sqrt{-g} \Big(  \nabla^\mu(\phi^2  \epsilon^\nu ) - \nabla^\nu(\phi^2  \epsilon^\mu )
-\phi^2\, \nabla^\mu\epsilon^\nu                                          \Big)\,,
\quad (\epsilon^\mu\in K.)\,,
$$
thus explicitating in this example the identical conservation of (\ref{conservlawkill}).

\subsection{Conformal Killing vectors}
\label{fixkillc}

Going back to (\ref{noethid3})
\beq 
-T^{\mu\nu} \nabla_{\!\mu}\, \epsilon_\nu + [{\cal L}]_{\!{}_A}\delta \phi^A + 
\nabla_{\!\sigma} \Big( T^{\sigma}_{\! \ub\ \mu} \epsilon^\mu  \Big) =0   \,,
\label{noethid3b}
\eeq
for arbitrary $\epsilon^\mu$, we realize that conformal Killing vectors,
$$
\nabla_{\!\mu}\, \epsilon_\nu+\nabla_{\!\nu}\, \epsilon_\mu = \lambda g_{\mu\nu}\,,
$$
are Noether symmetries of the truncated Lagrangian as long as the product $T \lambda$ is a divergence, 
with $T$ the trace of the Hilbert tensor. Thus the requirement is
$T \lambda
=\nabla_\mu D^\mu =\partial_\mu D^\mu$ for a vector density $D^\mu$. Then (\ref{noethid3b}) will be 
written as
\beq 
 [{\cal L}]_{\!{}_A}\delta \phi^A + 
\nabla_{\!\sigma} \Big( T^{\sigma}_{\! \ub\ \mu} \epsilon^\mu -\frac{1}{2} D^{\sigma} \Big) =0   \,,
\label{noethid3bconf}
\eeq

Condition $T \lambda =\partial_\mu D^\mu$
amounts\footnote{All our statements are local.} to require that the 
Euler-Lagrange derivatives of $T \lambda$ vanish, 
\beq
\frac{\delta (T \lambda)}{\delta \phi^A} = \frac{\delta T }{\delta \phi^A}\lambda 
- \frac{\partial T }{\partial\, \nabla_\mu\phi^A}\partial_\mu\lambda =0\,.
\label{check}
\eeq
This is a check that must be done in a case by case basis. 
In the particular case of homothetic Killing vectors ($\lambda=$ constant), the condition to implement
a Noether symmetry is that $T$ must be a divergence, that is $\frac{\delta T }{\delta \phi^A}=0$.

Obviously $T=0$ is a particular case of fulfillment of (\ref{check}). In such case, {\sl all} 
conformal Killing vectors yield Noether symmetries of the truncated theory. A well know example is 
Maxwell theory in a curved background -including the flat case-, where the Hilbert tensor satisfies
this condition in $d=4$ dimensions. Indeed for the Lagrangian 
$\displaystyle {\cal L} = -\frac{1}{4} \sqrt{-g}F^{\rho\sigma}F_{\rho\sigma}$
the Hilbert tensor is 
\beq
T^{\mu\nu}=
-g^{\mu\nu}{\cal L} + \sqrt{-g} F^{\mu}_{\ \sigma}F^{\sigma\nu}\,.
\label{hten}
\eeq
whose trace is $T= (4-d) {\cal L}$.

\section{Extended objects: the Nambu-Goto brane}
\setcounter{equation}{0}
\label{theNG}

Consider a Lorentzian manifold $P$ -the {\sl background}- and a submanifold $M$ -the 
{\sl p-brane, or brane}. The Levi Civita covariant derivative is defined on $P$. As notation, 
$x^\mu$ will be the coordinates on $P$ and $\sigma^a$ the coordinates on $M$. The embedding of the 
brane is locally defined by the functions $x^\mu(\sigma)$. Thus, on the tangent bundle $TM$,
$$
\frac{\pa }{\pa \sigma^a}=\frac{\pa x^\mu}{\pa \sigma^a}\frac{\pa }{\pa x^\mu} =: 
U^\mu_a \frac{\pa }{\pa x^\mu}.
$$
As a matter of language, we can also say that the {\sl world volume} $M$ is embedded in the 
{\sl target spacetime} $P$. 

Notice that $U^\mu_a$, defined with support on $M$, is a contravariant vector under 
reparametrizations of the background and a covariant vector under reparametrizations of the brane. 
\vs

Let $g_{\mu\nu}$ be the background metric. The induced metric on the brane is locally given by 
$$
\gamma_{ab}=g_{\mu\nu}U^\mu_a U^\nu_b\,.
$$
We consider massive branes. The p-brane is a p-dimensional spacelike object evolving along the 
time coordinate of the Lorentzian manifold $P$, thus the induced metric 
$\gamma_{ab}$ is again Lorentzian. Considering the index $a$ as a label, $U^\mu_a$ span a basis for 
$TM$. We can rise and lower indices with 
$\gamma_{ab},\, g_{\mu\nu}$, and their inverses, so for instance $U_\mu^a := \gamma^{ab}g_{\mu\nu}U^\nu_b$.
Note that 
\beq
t^\mu_{\ \nu} = U^\mu_a U_\nu^a
\label{proj-tan}
\eeq
is the projector from $TP|_{\!{}_M}$ to $TM$. Note also that given a vector $v^\mu$ in $TP|_{\!{}_M}$, 
its orthogonal projection to $TM$ is $v^a := U_\mu^a v^\mu$ (Proof: $t^\mu_{\ \nu} v^\nu \pa_\mu = 
U^\mu_a U_\nu^a v^\nu \pa_\mu= U_\nu^a v^\nu \pa_a =: v^a \pa_a $)\footnote{This result implies 
that for vectors already in $TM$ but expressed in the target spacetime coordinates, $U_\mu^a$ converts 
target spacetime indices into world volume indices.}. More geometric results are given 
in appendix \ref{apb}.

The Nambu-Goto (NG) Lagrangian of the p-brane is
\beq
{\cal L} = -\kappa\sqrt{-\gamma}\,,
\label{ng-lag}
\eeq
where $\gamma$ is the determinant of $\gamma_{ab}$ and $\kappa$ the tension of the brane. 
In the following we will take $\kappa=1$ for simplicity.

\vs

Regarding the analysis of the covariance  
of (\ref{ng-lag}) with respect to spacetime diffeomorphisms, there is the crucial difference as to 
whether we consider a dynamical or a fixed background. 
Of course if the background is dynamical the variational principle with Lagrangian 
(\ref{ng-lag}) is incomplete because the kinetic terms -for instance the Einstein-Hilbert Lagrangian- 
for the metric are missing. But the behaviour of (\ref{ng-lag}) under target spacetime diffeomorphisms can 
be studied nevertheless.

\subsection{ Dynamical background}
In the case of a brane we can think of it as a limit of a regular scalar field $\varphi$ when its 
support shrinks to $M$. If the field is peaked at a point with coordinates $x_0$, the actively 
transformed field $\varphi'$ will be peaked at coordinates $x'_0=x_0-\epsilon(x_0)$ so that 
$\varphi'(x')=\varphi(x)$. In the case of the brane this means that the embedding has changed from 
$x(\sigma)$ to $x'(\sigma)= x(\sigma)-\epsilon(x(\sigma))$. Alongside with it, the metric field will 
have changed under the active diffeomorphisms as well, with 
$g_{\mu\nu}(x)\to g'_{\mu\nu}(x)=g_{\mu\nu}(x)+{\cal L}_{\!{}_\epsilon} g_{\mu\nu}$. All in all 
the Lagrangian (\ref{ng-lag}) will have changed to ${\cal L}' = -\sqrt{-\gamma'}$, where 
$$
\gamma'_{ab}=g'_{\mu\nu}(x'(\sigma))\frac{\pa x'^\mu}{\pa \sigma^a}\frac{\pa x'^\nu}{\pa \sigma^b}\,.
$$
Notice in this expression above the subtle point that the active interpretation of diffeomorphisms 
imposes that the 
new spacetime metric $g'_{\mu\nu}$ is computed at the points $x'(\sigma)$ of the new embedding. 
Since the new embedding is $x'^\mu(\sigma)=x^\mu(\sigma)-\epsilon^\mu(x(\sigma))$, 
we obtain (to first order in the infinitesimal parameter)
\bea
\gamma'_{ab}&=&g'_{\mu\nu}(x(\sigma)-\epsilon(x(\sigma)))
\frac{\pa (x^\mu(\sigma)-\epsilon^\mu(x(\sigma))}{\pa \sigma^a}
\frac{\pa (x^\nu(\sigma)-\epsilon^\nu(x(\sigma))}{\pa \sigma^b}\nonumber\\
&=&(g'_{\mu\nu}(x)-{\cal L}_{\!{}_\epsilon}g_{\mu\nu})U^\mu_a U^\nu_b= 
g_{\mu\nu}(x)U^\mu_a U^\nu_b = \gamma_{ab}\,,
\eea
which in turn implies that, since the induced metric is invariant, $\gamma'_{ab}=\gamma_{ab}$, 
so it is indeed the 
Lagrangian: 
$\delta {\cal L}=0$. 
Note that the proof of invariance would be the same for a Lagrangian describing 
the minimal coupling of a $p$-brane with a $p+1$ form, which is a generalization of the coupling of 
the particle to the electromagnetic gauge field: again, the change of the world volume description 
of the brane will be matched by the active change of the $p+1$ form. 

Thus the requisite of general covariance is satisfied. Of course under world volume diffeomorphisms the 
Lagrangian is not invariant but behaves as a scalar density, see details in section \ref{Wv}. 
Summing up, 

\vs

\BOX{\vfill\vspace{1.5mm}\hspace{1.5mm}\it The Lagrangians of the NG brane and its generalizations are scalar densities under world vo\-lume 
diffeomorphisms and are invariant under target space diffeomorphisms.\vfill\vspace{1.5mm}}

\subsection{ Fixed background}
In the case of a fixed background, an active diffeomorphism will move the brane and all the other 
existing structures and fields according to their geometric character, {\it except for the metric}, 
which will remain unchanged. The reason is that in the active view we do not change the coordinates 
and hence a fixed background remains the same.

\vs
Now the Lagrangian will have changed to ${\cal L}' = -\sqrt{-\gamma'}$, where
$$
\gamma'_{ab}=g_{\mu\nu}(x'(\sigma))\frac{\pa x'^\mu}{\pa \sigma^a}\frac{\pa x'^\nu}{\pa \sigma^b}\,,
$$
(notice the crucial difference with the previous case: now we have written $g$ instead of $g'$)
and we obtain
$$
\gamma'_{ab}=(g_{\mu\nu}(x)-{\cal L}_{\!{}_\epsilon}g_{\mu\nu})U^\mu_a U^\nu_b\,,
$$
where ${\cal L}_{\!{}_\epsilon}$ is the Lie derivative under the vector field $\epsilon^\mu$. As for 
the determinant,
$$
\gamma'=\gamma (1- t^{\mu\nu}{\cal L}_{\!{}_\epsilon}g_{\mu\nu})\,.
$$
with $t^{\mu\nu}=\gamma^{ab}U_a^\mu U_b^\nu$ the projector defined above, (\ref{proj-tan}). Finally, 
for the variation of the Lagrangian we obtain 
\beq
\delta {\cal L}= {\cal L}'-{\cal L} = 
\frac{1}{2}\sqrt{-\gamma}\, t^{\mu\nu}{\cal L}_{\!{}_\epsilon}g_{\mu\nu}\,.
\label{deltaLb}
\eeq

Note that if $\epsilon$ is a Killing vector of the background, then 
${\cal L}_{\!{}_\epsilon}g_{\mu\nu}=0$, that is, the Lagrangian is invariant, but now under a 
rigid, not gauge, symmetry. Summing up, starting with a theory (\ref{ng-lag}) with invariance 
under the target spacetime diffeomorphisms, we have ended up, after freezing the background, 
with a theory in which the Killing symmetries of such
fixed background have morphed into rigid Noether symmetries of the new theory. This is the same phenomenon
already seen in field theory, Eq.(\ref{noethkill}), (\ref{killcurrent}).

We can look for the conserved current under this symmetry. Since 
($[{\cal L}]_{{}_\mu}:=\frac{\delta {\cal L}}{\delta x^\mu}$ is the Euler-Lagrange derivative)
\beq
\delta {\cal L}= [{\cal L}]_{{}_\mu} \delta x^\mu + 
\pa_a (\frac{\pa {\cal L}}{\pa U_a^\mu}\delta x^\mu)\,,
\label{standard}
\eeq
now for $\delta  x^\mu=-\epsilon^\mu$ being a Killing vector, we infer from (\ref{deltaLb}) that 
$\delta {\cal L}=0$ and the 
-on shell- conserved current is the -world volume- density vector (we factor out the infinitesimal 
parameter in $\epsilon^\mu$ so now $\epsilon^\mu$ will be finite)
\beq
J^a = \frac{\pa {\cal L}}{\pa U_a^\mu}\,\delta  x^\mu=
-\frac{\pa {\cal L}}{\pa U_a^\mu}\,\epsilon^\mu = \sqrt{-\gamma}\, U^a_\mu \epsilon^\mu=
\sqrt{-\gamma}\,\epsilon^a \,,\quad (\epsilon^\mu\in K.)\,.
\label{consc}
\eeq
This is a nice result: according to the intepretation of  $U^a_\mu$ made after Eq.(\ref{proj-tan}), 
we have found that

\vs

\BOX{\vfill\vspace{1.5mm}\hspace{1.5mm}\it For the NG brane, the conserved density current of a Noether symmetry generated by a 
Killing vector of the fixed background, 
is the densitized projection to $TM$ of such Killing vector.\vfill\vspace{1.5mm}}

\vs

Eventually there can be a conserved charge 
$\displaystyle G(\tau)= \int_{\tau} d^p\! \sigma\, J^0$, where the 
integration is at $\sigma^0\equiv \tau$ 
constant, but this conservation in $\tau$ will crucially depend on the spatial boundary of the brane. 
In the particular case of the massive particle in a fixed background, with Lagrangian 
$L=-m\sqrt{-g_{\mu\nu}\dot x^\mu\dot x^\nu}$, this conserved charge is 
$\displaystyle G= \frac{1}{\sqrt{-g_{\mu\nu}\dot x^\mu\dot x^\nu}}\epsilon^\mu g_{\mu\nu}\dot x^\nu$.

But regardless of being conserved 
or not, $G(\tau)$ is nevertheless the gauge generator in the canonical formalism. Since from 
(\ref{consc}) $\displaystyle J^0= \frac{\pa {\cal L}}{\pa U_0^\mu}\,\delta  x^\mu = - p_\mu \epsilon^\mu $, 
with $p_\mu$ the momenta canonically conjugate to 
the target spacetime coordinates, we obtain the trivial result in the canonical formalism, for the 
variation of the embedding, 
$$
\delta   x^\mu=\{x^\mu,\,G\}_{\!{}_{{\rm equal}\, \tau}} = 
-\{x^\mu,\,\int_{\tau} d^p\! \sigma\, p_\rho\, \epsilon^\rho\}_{\!{}_{{\rm equal}\, \tau}}=-\epsilon^\mu
$$
\vs

The algebra of the Noether symmetries associated with Killing symmetries of the fixed background 
reproduces that of the Killing symmetries. If $\vec\epsilon$ and  $\vec\eta$ are two Killing vectors, 
then  $G_{\!{}_{\vec\epsilon}}$ and $G_{\!{}_{\vec\eta}}$ satisfy 
$\{G_{\!{}_{\vec\epsilon}},\,G_{\!{}_{\vec\eta}}\} = G_{\!{[\vec \eta,\,\vec\epsilon]}}$\,.

\vs

Again with the massive particle, consider a stationary background
(with $x_0$ the background time coordinate), so that $\pa_{x_0}$ is a Killing vector (
$\epsilon^\mu = \delta^\mu_0$). If $\tau$ is the world line parameter, the induced metric has a 
single component $\gamma_{\tau\tau}=\dot x^2 =\gamma$, where $\dot x^2 =
g_{\mu\nu} \dot x^\mu\dot x^\nu$. 
The embedding is $U^\mu_\tau = \dot x^\mu$ and $U_\mu^\tau= 
\gamma^{-1}g_{\mu\nu} U^\nu_\tau = g_{\mu\nu}\frac{\dot x^\nu}{\dot x^2}$, so  
$\sqrt{-\gamma}\,U_\mu^\tau=
-g_{\mu\nu}\frac{\dot x^\nu}{\sqrt{-\dot x^2}} =: - p_\mu$. Applying (\ref{consc}), the 
conserved quantity $J^\tau\equiv G$ is $G = -\delta^\mu_0 p_\mu= -p_0$, which is the expected 
result for a mechanical system with a cyclic coordinate $x_0$, namely that its conjugate momentum 
$p_0$ is a constant of motion.

\section{Noether identities for the brane}
\setcounter{equation}{0}
\label{NIb}

It is instructive, regardless of the Killing condition, to compute directly the divergence of 
$\epsilon^a:= U^a_\mu \epsilon^\mu$ for an arbitrary, infintesimal, $\epsilon^\mu$. We have
\bea
\frac{1}{\sqrt{-\gamma}}\pa_a (\sqrt{-\gamma}\,\epsilon^a) &=& \nabla_{\!a} \epsilon^a = 
\nabla_{\!a}(U^a_\mu \epsilon^\mu) = \tilde\nabla_{\!a}(U^a_\mu) \epsilon^\mu+ U^a_\mu\nabla_{\!a}
( \epsilon^\mu)\nonumber\\
&=& \gamma^{ab}g_{\mu\nu}\tilde\nabla_{\!a}(U_b^\nu) \epsilon^\mu + U^a_\mu U_a^\rho g^{\mu\nu} 
\bar\nabla_{\!\rho}\epsilon_\nu= \gamma^{ab}g_{\mu\nu}K_{ab}^\nu\epsilon^\mu+ \frac{1}{2}t^{\rho\nu} 
{\cal L}_{\!{}_\epsilon}g_{\rho\nu}\,,
\label{standard2}
\eea
(where we have used that $\tilde\nabla_{\!a} U_b^\nu=K_{ab}^\nu$ is the extrinsic curvature of the 
brane. See the Appendix \ref{apb} for details and also for a discussion on the covariant derivatives 
$\nabla_{\!a}, \tilde\nabla_{\!a},\bar\nabla_{\!\rho}$). Note that (\ref{standard2}) can be 
rearranged as ($K^\nu$ is the trace of the extrinsic curvature)
\beq
\frac{1}{2}\sqrt{-\gamma}\,t^{\rho\nu} {\cal L}_{\!{}_\epsilon}g_{\rho\nu}= 
-\sqrt{-\gamma}\,g_{\mu\nu}K^\nu\epsilon^\mu+ \pa_a (\sqrt{-\gamma}\,\epsilon^a)\,,
\label{standard3}
\eeq
which is exactly (\ref{standard}) applied to our case with $\delta x^\mu = -\epsilon^\mu$
and $\epsilon^a:= U^a_\mu \epsilon^\mu$. 
Eq.(\ref{standard3}) is an identity -actually, it is a Noether identity, see next subsection- 
with $\epsilon^\mu$ arbitrary. As a byproduct we obtain, comparing 
(\ref{standard}) and (\ref{standard3}),
\beq
[{\cal L}]_{{}_\mu}:=\frac{\delta {\cal L}}{\delta x^\mu}= \sqrt{-\gamma}\,g_{\mu\nu}K^\nu\,,
\label{eomNG}
\eeq
that is, we have obtained the well known result (see for instance \cite{Carter:2000wv}) that the EOM for 
the NG Lagrangian is the 
vanishing of the trace of the extrinsic curvature, with no need to make the explicit computation of the EOM.
 
\subsection{Target spacetime Noether identities}

As anticipated above, Eq.(\ref{standard3}) has yet another interpretation: it is the {\sl Noether 
identity for the target spacetime diffeomorphism invariance of the NG Lagrangian (\ref{ng-lag}) when 
the background is dyna\-mical}. Indeed, given some fields -or field components- $\psi^A$ and expressing 
the active infinitesimal gauge transformation as 
(\ref{gaugetr}) (with $\delta x^\mu(\sigma) = -\epsilon^\mu(x(\sigma))$) 
but now for the world volume action,
\beq
\delta \psi^A = R^A_\rho \epsilon^\rho + Q^{A\,a}_\rho \partial_a\epsilon^\rho\,,
\label{gaugetra}
\eeq
where $\psi^A$ is a generic field or field component and 
$\partial_a\epsilon^\rho = U^\mu_a \partial_\mu\epsilon^\rho$.
The Noether identity -in the version with the arbitrary functions in it- 
takes the form (\ref{noethid1}), but with derivatives with respect to the 
$\sigma^a$ coordinates of the brane,
\beq
[{\cal L}]_A \delta\psi^A-\partial_a([{\cal L}]_A Q^{A\,a}_\rho\epsilon^\rho)=0\,,
\label{noethid1rev}
\eeq
In our case the fields $\psi^A$ are the target coordinates $x^\mu(\sigma)$ and the metric 
$g_{\mu\nu}(x(\sigma))$. For the metric $\delta g_{\mu\nu}= {\cal L}_{\!{}_\epsilon}g_{\mu\nu}$, 
which implies (with $A=(\mu\nu)$), for the world volume description, 
\beq
 \delta g_{\mu\nu} = R_{(\mu\nu)\rho}\epsilon^\rho +Q_{(\mu\nu)\rho}^a \partial_a\epsilon^\rho\,,\ 
 {\rm with}\,\
R_{(\mu\nu)\rho}= \pa_\rho g_{\mu\nu},\quad Q_{(\mu\nu)\rho}^a = g_{\rho\nu}U^a_\mu + g_{\mu\rho}U^a_\nu\,,
\label{deltag}
\eeq
then, taking this into account, together with $\delta x^\mu = -\epsilon^\mu$, we obtain the Noether
identity associated with target spacetime diffeomorphism invariance of the NG brane, 
($\displaystyle [{\cal L}]^{\mu\nu}= \frac{\partial {\cal L}}{\partial g_{\mu\nu}},\ [{\cal L}]_{\rho}
= \frac{\partial {\cal L}}{\partial x^{\rho}}$)
\beq
 [{\cal L}]^{\mu\nu} \partial_\rho g_{\mu\nu}-[{\cal L}]_{\rho} -
 \partial_a \Big([{\cal L}]^{\mu\nu}( g_{\rho\nu}U^a_\mu + g_{\mu\rho}U^a_\nu)\Big)=0\,,
\label{noethid1rev2}
\eeq
identically. Using the fact that 
$\displaystyle\frac{\delta {\cal L}}{\delta g_{\mu\nu}}= 
-\frac{1}{2}\sqrt{-\gamma}\,t^{\mu\nu}$ and Eq.(\ref{eomNG}), one can check that 
Eq.(\ref{noethid1rev2}), saturated with $\epsilon^\rho$, becomes (\ref{standard3}). 

\vs

The advantage of the Noether identity formulation, Eq.(\ref{noethid1rev})-(\ref{noethid1rev2}), 
is that it shows how the 
me\-ca\-nism works beyond the NG Lagrangian, that is, when we consider corrections to it. 
In the first term of Eq.(\ref{noethid1rev}) there will always be the term 
$\frac{\delta {\cal L}}{\delta g_{\mu\nu}}\delta g_{\mu\nu}$. If the background is fixed the 
factor $\frac{\delta {\cal L}}{\delta g_{\mu\nu}}$ will not vanish in general but ${\delta g_{\mu\nu}}$
will vanish -by definition- for the variations engendered by the Killing vectors of the metric, 
thus defining the on shell conserved current, see Eq.(\ref{noethid1rev}), (\ref{deltag}), 
$J^a=-\frac{\delta {\cal L}}{\delta g_{\mu\nu}} Q_{(\mu\nu)\rho}^a\epsilon^\rho$.

\vs


\subsection{World volume Noether identities}
\label{Wv}
Clearly, the NG Lagrangian (\ref{ng-lag}) is generally covariant with respect to world volume 
diffeomorphisms, so that in the active intepretation, with 
$\delta \sigma^a = -\eta^a(\sigma)$, we will have $\delta x^\mu = \eta^a\partial_a x^\mu(\sigma)$ and, 
due to the variation $\delta x^\mu$, there will be a vicarious variation of $g_{\mu\nu}(x(\sigma))$, 
but the metric won't have any variation by its own, unlike the case of the target spacetime 
diffeomorphisms (both the target coordinates and the metric components are scalars 
under world volume diffeomorphisms). All in all we find the world volume scalar density behaviour 
$\delta{\cal L}=\partial_a(\eta^a{\cal L})$. The Noether identity becomes
$$
[{\cal L}]_{{}_\nu}\,  U_a^\nu =0\,,
$$
identically. Its contents for the NG case is just $K^\mu g_{\mu\nu}U_a^\nu =0$, which is an identity 
proved directly in Appendix \ref{apb}, see Eq.(\ref{ortho}).

\section{Conclusions}
\label{concl}

In this paper we have extended the results in \cite{Pons:2009nb} to curved spacetime and in addition 
we have made a parallel analysis for the case of extended objects embedded in the background. We have shown
in particular the relationship between the canonical tensor, Belinfante energy-momentum tensor, 
and Hilbert energy-momentum tensor. 
When these results are applied lo matter Lagrangians truncated to a fixed background, we have shown that
the Killing symmetries of the background become ordinary, non-gauge, Noether symmetries of the 
truncated theory. We stress the role of the Belinfante tensor in the construction both of the Noether 
conserved current and of the canonical generator of the symmetries.

We have followed similar lines to analyze the case of NG branes embedded in the background, distinguishing
the two types of gauge symmetries available, namely target space diffeomorphisms and world volume 
diffeomorphisms. We use the Noether identities of the former type to give an alternative 
derivation of the EOM of the theory truncated to a fixed background. 
Again, the Killing symmetries of the fixed background give rise to Noether symmetries of the world 
volume theory, with a simple geometric intepretation of the conserved currents.

\begin{appendix}

\section{Poincar\'e symmetries as the Killing symmetries for Maxwell theory in flat spacetime.}
\setcounter{equation}{0}
\label{apa}
In this appendix we do not intend to give a complete analysis on how 
the Poincar\'e transformations for the Maxwell gauge field in flat spacetime can 
be reproduced within the canonical formalism. 
Our much more modest goal here is to underline the crucial relevance of the Belinfante tensor 
in such an endeavour. 

In flat spacetime (with mostly plus metric $\eta_{\mu\nu} = (-, +, +, +)$) 
the Maxwell Lagrangian is 
${\cal L} = -\frac{1}{4} F^{\mu\nu}F_{\mu\nu}$. The momentum canonically conjugate to $A_\mu$ is
$\displaystyle\pi^\mu =\frac{\partial {\cal L}}{\partial \dot A_\mu} = -F^{0\mu}$, 
which identifies the primary constraint $\pi^0\simeq 0$ (``$\simeq$'' means equality on shell). 
Hilbert energy momentum tensor $T^{\mu\nu}$, (\ref{hten}), is obtained by the standard methods.
Belinfante energy momentum tensor is, from (\ref{belinf}),
$$
T^{\mu\nu}_{\! \ub} = T^{\mu\nu}- 
[{\cal L}]_{\!{}_A} Q^{A\,\mu}_{\quad\, \sigma}\eta^{\sigma\nu}=
-\eta^{\mu\nu}{\cal L} + F^{\mu}_{\ \sigma}F^{\sigma\nu}-
\partial_\rho(F^{\rho \mu})A_\sigma\eta^{\sigma\nu}\,.
$$
Notice that the last term, which vanishes on shell, distinguishes this tensor 
from Hilbert tensor $T^{\mu\nu}$.

Acording to (\ref{noethkill}), the conserved current is 
$$
J^\mu = T^{\mu}_{\! \ub\ \nu}\,\epsilon^\nu = -{\cal L}\,\epsilon^\mu+ F^{\mu}_{\ \sigma}F^{\sigma}_{\ \nu}
\epsilon^\nu-\partial_\rho(F^{\rho \mu}) A_\nu\epsilon^\nu
$$
with the Killing vector $\epsilon^\nu = a^\nu + \omega^\nu_{\ \rho}\, x^\rho$ 
describing the infinitesimal Poincar\'e transformations. 

The candidate to be the generator of these transformations is $G =\int d^3 x \,J^0$, 
which expressed with the canonical variables is 
($\displaystyle\mu =0,i)$, with $B^i = \frac{1}{2}\epsilon^{i j k} F_{jk}$,
\beq
G(\epsilon^\mu)= \int d^3 x \Big( \frac{1}{2} (\vec \pi^2 +\vec B^2)\, \epsilon^0 + 
(\vec \pi\times\vec B)\cdot\vec \epsilon- (\partial_ i \pi^i) A_\mu\epsilon^\mu\Big)
\label{theG}
\eeq
There is an ambiguity in $G$ because a term linear in the primary constraint $\pi^0$ can be added 
without consequences in tangent space -the pullback  
of this term from phase space to tengent space vanishes identically-. This ambiguity can be solved
in the light of the analysis in \cite{Pons:1999az} but we will not dwell further on this issue.

Although it is not our goal here, let us mention that, following the results in \cite{Pons:1999az}, 
one can immediately identify a problem with 
this expression (\ref{theG}), in what concerns its eventual role as generator of Noether symmetries. 
As a matter of fact,
$$
\{G,\, \pi^0\} = -(\partial_ i \pi^i) \epsilon^0\,,
$$
where one can identify in the right hand side the secondary constraint of Maxwell theory 
-the Gauss law-  $\partial_ i \pi^i\simeq 0$. The presence of a -necessarily first class- 
secondary constraint in the rhs of the computation of the Poisson bracket of $G$ with a -first class- 
primary constraint signals an obstruction for the projectability -from tangent space to phase space- 
of some of the transformations generated by $G$.In our case we infer, from the results of \cite{Pons:1999az} 
section III B, that this quantity $G$ won't generate the time translation for $A_0$. 
We say that this transformation, which can be written 
straightforwardly in tangent space, $\delta A_0= \delta t \dot A_0$, is {\sl not} projectable to phase space. 
It can not be retrieved by canonical methods alone\footnote{A parallel situation takes place
in the canonical formalism of Einstein-Hilbert Lagrangian for general relativity, where some 
diffeomorphisms are not projectable to phase space, see \cite{Pons:1996av,Pons:2003uc} for an analysis 
and a way out of that situation.} 
because, owing to the fact that $\pi^0$ is a constraint, the velocity $\dot A_0$ can not be 
expressed in terms of the canonical variables. More details can be found in \cite{Pons:1999az}.

But let us compute some projectable transformations. All transformations for the space components 
$A_i$ of the gauge field are projectable and are not affected by the ambiguity in $G$ mentioned above.
So we have
$$
\delta_\epsilon A_i = \{A_i,\, G(\epsilon^\mu)\} = \pi^i \epsilon^0 + (\vec B\times \vec\epsilon )^i + 
\partial_i (A_\nu\epsilon^\nu).
$$
It is immediate to verify that the pullback of this transformation to tangent space is 
exactly what is expected. Indeed, after implementing the pullback $\pi^i \to F_{0 i}$,
\beq
\delta_\epsilon A_i =  F_{0 i} \epsilon^0 -F_{ij} \epsilon^j+ (\partial_i A_\nu) \epsilon^\nu 
+ A_\nu\partial_i\epsilon^\nu= \epsilon^\nu\partial_\nu A_i+ A_\nu\partial_i\epsilon^\nu=
{\cal L}_\epsilon A_i\,,
\label{deltalie}
\eeq
with ${\cal L}_\epsilon$ the Lie derivative under the -Poincar\'e- Killing vector 
$\epsilon^\nu = a^\nu + \omega^\nu_{\ \rho}\, x^\rho$. In fact, reminiscent of the original 
diffeomorphism covariance -before truncation to a fixed background- of our matter Lagrangian, 
Eq.(\ref{deltalie}) gives indeed the right transformation under general diffeomorphisms 
(arbitrary $\epsilon^\mu$) but only for Poincar\'e transformations we do 
have a Noether symmetry in flat spacetime.

\vs

Let us stress the crucial role played in this computation by the term 
$\, (\partial_ i \pi^i) A_\mu\epsilon^\mu\, $ in (\ref{theG}), which would have been missing should 
we had used the Hibert tensor instead of Belinfante's. The projectability problem mentioned above
is solved by adding to the generator (\ref{theG}) the generator of a particular $U(1)$ gauge transformation. 
Details in the more general framework of Einstein-Yang-Mills theories are given in \cite{Pons:1999xu}.
\section{Basic notions on the extrinsic curvature}
\setcounter{equation}{0}
\label{apb}
Consider a Riemannian or Lorentzian manifold $P$ -the {\sl background}- and a submanifold $M$ -the 
{\sl brane}. The Levi Civita covariant derivative is defined on $P$. As notation, $x^\mu$ will be the 
coordinates on $P$ and $\sigma^a$ the coordinates on $M$. The embedding of the brane is locally defined 
by the functions $x^\mu(\sigma)$. Thus, on $TM$ (tangent space)
$$
\frac{\pa }{\pa \sigma^a}=\frac{\pa x^\mu}{\pa \sigma^a}\frac{\pa }{\pa x^\mu} =: U^\mu_a \frac{\pa }
{\pa x^\mu}.
$$
Notice that $U^\mu_a$, defined with support on $M$, is a contravariant vector under 
reparametrizations of the background and a covariant vector under reparametrizations of the brane. 
\vs

Let $g_{\mu\nu}$ be the background metric. The induced metric on the brane is locally given by 
\beq
\gamma_{ab}=g_{\mu\nu}U^\mu_a U^\nu_b\,.
\label{ind-metr}
\eeq

Considering the index $a$ as a label, $U^\mu_a$ span a basis for $TM$. We can enlarge this basis 
to form a basis for $TP|_{\!{}_M}$ ($TP|_{\!{}_M}= TM \oplus T^\perp\! M$) 
with some new vectors $U^\mu_{a'}$, 
with labels $a'$, which we will take orthonormal and orthogonal to $TM$, that is, with the notation 
$\ua =(a,a')$, 
\beq g_{\mu\nu}U^\mu_{\ua}U^\nu_{a'} = \eta_{\ua a'}.
\label{enlarg}
\eeq  
We can define the matrix $U^\ua_\mu$ as the inverse matrix to $U^\mu_\ua$. Notice that these 
inverses still go through the rule of raising and lowering indices with the corresponding metric 
or inverse metric: 
$U^a_\mu= \gamma^{ab}g_{\mu\nu}U^\nu_b\,,\ U^{a'}_\mu= \eta^{a'b'}g_{\mu\nu}U^\nu_{b'}$.
\vs

Consider vector fields ${\bf X},{\bf Y}$ in $TM$. One can compute $\bar\nabla_{\!\bf X} {\bf Y}$ 
under the background Levi Civita covariant derivative $\bar\nabla$ and decompose it in its tangent 
part in $TM$ and an orthogonal part in $T^\perp\! M$. In fact we have 
\beq\delta^\mu_{\ \nu} = U^\mu_a U^a_\nu + U^\mu_{a'} U^{a'}_\nu =:t^\mu_{\ \nu} + h^\mu_{\ \nu}\,,
\label{decomp}
\eeq 
where $t^\mu_{\ \nu}$ projects onto $TM$ and $h^\mu_{\ \nu}$ onto $T^\perp\! M$. The projection to $TM$ will 
define the {\sl induced covariant derivative} on $TM$ and the projection to $T^\perp\! M$ will define 
the {\sl extrinsic curvature}.
\vs
\subsection{Induced covariant derivative}
With ${\bf X}= X^a\pa_a = X^\mu \pa_\mu$ (that is, $X^\mu=X^a U^\mu_a$) and same for ${\bf Y}$, we 
obtain, for the tangent part,
$$
(X^\mu\bar\nabla_{\!\mu} Y^\nu)t^\lambda_{\ \nu} \pa_\lambda= X^a\Big(\pa_a Y^c + (\pa_a U_b^\rho + 
U^\mu_a U^\nu_b\bar\Gamma_ {\mu\nu}^\rho)U^c_\rho Y^b\Big)\pa_c\,,
$$
which defines the induced covariant derivative $(X^a\nabla_{\!a} Y^c) \pa_c$ with connection
\beq
\Gamma_{ab}^c := (\pa_{ab} x^\rho + U^\mu_a U^\nu_b \bar\Gamma_ {\mu\nu}^\rho)U^c_\rho\,.
\label{ind-con}
\eeq
This induced connection is obviously torsionless. One can easily check that it is metric compatible 
by taking for instance adapted coordinates. Thus we infer that $\Gamma_{ab}^c$ is the Levi Civita 
connection for $M$.
\subsection{Extrinsic curvature}
The $T^\perp\! M$ component of $\bar\nabla_{\!\bf X} {\bf Y}$ will define the second fundamental 
form or extrinsic curvature, $K^\mu ({\bf X},{\bf Y})= (\bar\nabla_{\!X} Y^\nu)h^\mu_\nu $. Its 
components in the basis $U_ {a'}^\mu$ of $T^\perp\! M$ turn out to be
\beq
K^{c'}_{ab}=(\pa_{ab} x^\rho + U^\mu_a U^\nu_b \bar\Gamma_{\mu\nu}^\rho)U^{c'}_\rho\,,
\label{extr}
\eeq
which is obviously symmetric. Note that from (\ref{ind-con}) and (\ref{extr}) we can write
\beq
\pa_{ab} x^\rho + U^\mu_a U^\nu_b \bar\Gamma_ {\mu\nu}^\rho = \Gamma_{ab}^c U_c^\rho + 
K^{c'}_{ab}U_{c'}^\rho\,.
\label{compl}
\eeq

Now let us remember what we said on the double vector character of $U_a^\mu$ under background and 
brane reparametrizations. To take into account both behaviours, the complete covariant derivative 
$\tilde\nabla_{\!a}$ is defined so that
\beq
\tilde\nabla_{\!a} U_b^\rho = \pa_a U_b^\rho + U^\mu_a \bar\Gamma_ {\mu\nu}^\rho U^\nu_b - 
\Gamma_{ab}^c  U_c^\rho\,,
\label{compl2}
\eeq
which, in view of (\ref{compl}), can be writen as
\beq
\tilde\nabla_{\!a} U_b^\rho =K^{c'}_{ab}U_{c'}^\rho =: K^\rho_{ab}\,.
\label{extr2}
\eeq

\vs

Now we can directly prove that $K^\rho_{ab}$ is orthogonal to $TM$. Let us compute, using (\ref{extr2}),
\bea
A_{abc}&=& K^\mu_{ab}g_{\mu\nu}U^\nu_c =(\tilde\nabla_{\!a} U_b^\mu) g_{\mu\nu}U^\nu_c = 
\tilde\nabla_{\!a}(U_b^\mu g_{\mu\nu}U^\nu_c)-U_b^\mu g_{\mu\nu}(\tilde\nabla_{\!a} U^\nu_c) \nonumber \\ 
&=& \nabla_{\!a}(\gamma_{bc})- (\tilde\nabla_{\!a} U^\nu_c) g_{\mu\nu}U_b^\mu =-A_{acb}\,,
\label{ortho}
\eea
but it is well known that a three index quantity $A_{abc}$ symmetric in two indices ($a,b$) 
and antisymmetric in two indices [$b,c$] must vanish.
\subsection{Expressions for the extrinsic curvature}
Since $U^\mu_{b'}$ are taken orthonormal, we have
\bea
K^{c'}_{ab} &=& \eta^{c'd'} K^\mu_{ab}g_{\mu\nu}U^\nu_{d'}= \eta^{c'd'}
(\tilde\nabla_{\!a} U_b^\mu)g_{\mu\nu}U^\nu_{d'}= -\eta^{c'd'}U_b^\mu g_{\mu\nu}
\bar\nabla_{\!a} U^\nu_{d'}= -\eta^{c'd'}U_b^\mu g_{\mu\nu}U^\rho_a
\bar\nabla_{\!\rho} U^\nu_{d'}\nonumber\\
&=& U_b^\mu g_{\mu\nu}\bar\nabla_{\!a} U^{\nu c'}\,,
\label{wehave}
\eea
or
\beq
(\bar\nabla_{\!a} U^{\nu c'}) t_\nu^\rho = K^{c'}_{ab} \gamma^{bc}
U^{\rho}_c\,.
\label{wehave2}\eeq

In the particular case of a spacelike codimension one brane on a Lorentzian background, then 
$c'=0,\ \eta^{00}=-1\ {\rm (mostly\ plus)},\ U^\mu_{c'}=: n^{\mu}$ (notice from (\ref{enlarg}) 
that $n_\mu n^\mu =-1$),$\ K^{c'}_{ab}=:  K_{ab}$, 
which is of interest for the ADM formalism of general relativity \cite{Arnowitt:1962hi}, we will have, 
from (\ref{wehave2}) and using $n_\mu\bar\nabla_{\!a} n^\mu=0$, 
$\bar\nabla_{\!a} n^\mu = {K_a}^b U^\mu_b$.
\vs

If we take an adapted coordinatization, so that $\sigma^a = x^i$ (a subset of the background 
coordinates) and $M$ is locally defined by the remaining background coordinates $x^{i'}$ as 
functions of $x^{i}$, we will have $U_a^\mu = \delta_i^\mu$ etc., so, from (\ref{wehave}),
$$ 
K^{c'}_{ij}=-\eta^{c'd'}g_{j\nu}\bar\nabla_{\!i} U^\nu_{d'} = -\eta^{c'd'}\bar\nabla_{\!i}  U_{j d'}= 
-\eta^{c'd'}\bar\Gamma_{ij}^\rho U_{\rho d'} = -\eta^{c'd'}\bar\Gamma_{ij}^{j'} U_{j' d'}\,,
$$
where $U_{j d'}=0$ has been used in the last two equalities\,. In the ADM case $M$ will be an equal 
time surface and we will have $ K_{ij}=\bar\nabla_{\!i} n_j = \bar\Gamma_{ij}^0 n_0$.
\vs

Continuing with adapted coordinates, the $i,j$ components of the Lie derivative of the background 
metric read
$$
{\cal L}_{\!{}_{U_{\!{}_{c'}}}}g_{ij} = U_{c'}^\mu \bar \nabla_{\!\mu} g_{ij}- 
g_{\mu j}\bar \nabla_{\!i} U_{c'}^\mu - g_{\mu i}\bar \nabla_{\!j} U_{c'}^\mu = 
-2 \bar \nabla_{\!i} U_{j c'} = 2\, \eta_{c'd'}K^{d'}_{ij}\,,
$$
so 
$$
K^{c'}_{ij}= \frac{1}{2}\eta^{c'd'}{\cal L}_{\!{}_{U_{\!{}_{d'}}}}g_{ij}\,,
$$
which in the ADM case yields $K_{ij}= -\frac{1}{2}{\cal L}_{\!n}g_{ij}$.
\vs
\subsection{Dynamical intepretation of the extrinsic curvature}
Here we show that the vanishing of the trace of the extrinsic curvature is just the contents of the 
EOM of the NG brane.

First let us write the trace of the extrinsic curvature, from (\ref{extr2})
$$
K^\rho= \gamma^{ab}K^\rho_{ab}= \gamma^{ab}\tilde\nabla_{\!a} U_b^\rho = \Box x^\rho + 
t^{\mu\nu}\bar\Gamma_ {\mu\nu}^\rho ,
$$
where $\Box$ is the Dalembertian of the world volume (the brane) with $x^\rho$ taken, as it is, 
as a scalar under world volume diffeomorphisms, and $t^{\mu\nu}=\gamma^{ab}U^\mu_a U^\nu_b$ is the 
projector onto $TM$ defined in (\ref{decomp}).

\vs

Now consider the NG dynamics, defined by the Lagrangian ${\cal L} = -\kappa\sqrt{-\gamma}$, with 
$\gamma$ the determinant of $\gamma_ {ab}$. The E-L functional derivatives turn out to be
$$
\frac{\delta {\cal L}}{\delta x^\mu}= \kappa\, g_{\mu\nu} \sqrt{-\gamma}K^\nu\,.
$$
Thus the contents of the EOM is the vanishing of the trace of the extrinsic curvature -the curvature 
vector \cite{Carter:2000wv}. 
For the $0$-brane (particle), $K^\mu=0$ is the geodesic equation with arbitrary time parameter. In fact 
$$
K^\rho_{particle} = 
{\mathcal P}_\sigma^{\ \rho}(\ddot x^\sigma + \dot x^\mu \dot x^\nu \Gamma_{\mu\nu}^\sigma)\,,
\ {\rm with}\ {\mathcal P}_\sigma^{\ \rho}= 
(\delta_\sigma^{\ \rho} - g_{\sigma\lambda}\frac{\dot x^\lambda \dot x^\rho}{\gamma})\,,
$$
where ${\mathcal P}_\sigma^{\ \rho}$ is the projector orthogonal to the velocity $\dot x^\mu$. 
So $K^\mu=0$ means that 
$\ddot x^\sigma + \dot x^\mu \dot x^\nu \Gamma_{\mu\nu}^\sigma = \alpha\, \dot x^\sigma$ 
for some function $\alpha$. This function  
$\alpha$ vanishes when $\tau$ is taken to be the proper time, or a parameter proportional to it. 

\end{appendix}


\begin{thebibliography}{99}
\bibitem{Pons:2009nb}
  J.~M.~Pons,
  ``Noether symmetries, energy-momentum tensors and conformal invariance in classical field theory,''
  J.\ Math.\ Phys.\  {\bf 52} (2011) 012904
  doi:10.1063/1.3532941
  [arXiv:0902.4871 [hep-th]].

  
\bibitem{Belinfante}
 F. ~J. ~Belinfante, ``On the spin angular momentum of mesons'',  Physica  6,  (1939). 887--898.
 
 
\bibitem{Noether:1918zz}
  E.~Noether,
  ``Invariant Variation Problems,''
  Gott.\ Nachr.\  {\bf 1918} (1918) 235
   [Transp.\ Theory Statist.\ Phys.\  {\bf 1} (1971) 186]
  doi:10.1080/00411457108231446
  [physics/0503066].

\bibitem{Hilbert}
D.~Hilbert, ``Die Grundlagen der Physik",  Nachr. Ges. Wiss.
G\"{o}ttingen. {\bf 27} (1915), 395-407.

\bibitem{Rosenfeld}
L.~Rosenfeld, ``Sur le tenseur d'impulsion-\'energie",  
M\'em. Acad. Roy. Belg. Sci. {\bf 18} (1940) 1-30.

\bibitem{Wald:1984rg}
  R.~M.~Wald,
  ``General Relativity,''
  Chicago, Usa: Univ. Pr. ( 1984) 491p
  doi:10.7208/chicago/9780226870373.001.0001
  
\bibitem{Pons:1999az}
  J.~M.~Pons and J.~A.~Garcia,
  ``Rigid and gauge Noether symmetries for constrained systems,''
  Int.\ J.\ Mod.\ Phys.\ A {\bf 15} (2000) 4681
  doi:10.1142/S0217751X00001968, 10.1142/S0217751X00001966
  [hep-th/9908151].
  
\bibitem{Pons:1996av}
  J.~M.~Pons, D.~C.~Salisbury and L.~C.~Shepley,
  ``Gauge transformations in the Lagrangian and Hamiltonian formalisms of generally covariant theories,''
  Phys.\ Rev.\ D {\bf 55} (1997) 658
  doi:10.1103/PhysRevD.55.658
  [gr-qc/9612037].
  
\bibitem{Pons:2003uc}
  J.~M.~Pons,
  ``Generally covariant theories: The Noether obstruction for realizing certain space-time diffeomorphisms in phase space,''
  Class.\ Quant.\ Grav.\  {\bf 20} (2003) 3279
  doi:10.1088/0264-9381/20/15/301
  [gr-qc/0306035].
  36;
   
\bibitem{Pons:1999xu}
  J.~M.~Pons, D.~C.~Salisbury and L.~C.~Shepley,
  ``Gauge transformations in Einstein-Yang-Mills theories,''
  J.\ Math.\ Phys.\  {\bf 41} (2000) 5557
  doi:10.1063/1.533425
  [gr-qc/9912086].
  
\bibitem{Arnowitt:1962hi}
  R.~L.~Arnowitt, S.~Deser and C.~W.~Misner,
  Chapter 7 (pp. 227–265) of Louis Witten (ed.), 
  Gravitation: An introduction to current research, Wiley: New York, 1962.
  Republished: ``The Dynamics of general relativity,''
  Gen.\ Rel.\ Grav.\  {\bf 40} (2008) 1997
  doi:10.1007/s10714-008-0661-1
  [gr-qc/0405109].
  
\bibitem{Carter:2000wv}
  B.~Carter,
  ``Essentials of classical brane dynamics,''
  Int.\ J.\ Theor.\ Phys.\  {\bf 40} (2001) 2099
  doi:10.1023/A:1012934901706
  [gr-qc/0012036].

\end{thebibliography}
\end{document}